\documentclass[conference]{IEEEtran}
\usepackage[T1]{fontenc}
\usepackage[utf8]{inputenc}
\usepackage{cite}
\usepackage{url}
\usepackage{array}
\usepackage{booktabs}
\usepackage{tabularx}
\usepackage{amsmath}
\usepackage{enumitem}
\usepackage{multirow}
\usepackage{xcolor}
\usepackage{balance}
\usepackage{graphicx}
\usepackage{tikz}
\usetikzlibrary{arrows.meta,positioning}
\usepackage{microtype}
\usepackage[hidelinks]{hyperref}

\title{Beyond Task Success: An Evidence-Synthesis Framework for Evaluating, Governing, and Orchestrating Agentic AI}

\author{%
\IEEEauthorblockN{Christopher Koch}
\IEEEauthorblockA{Independent Researcher}
\and
\IEEEauthorblockN{Joshua A. Wellbrock}
\IEEEauthorblockA{Independent Researcher}
}

\newcommand{\odta}{ODTA}
\newcommand{\maeb}{MAEB}

\begin{document}
\maketitle

\begin{abstract}
Agentic AI systems plan, use tools, maintain state, and act across multi-step workflows with external effects, meaning trustworthy deployment can no longer be judged by task completion alone. The current literature remains fragmented across benchmark-centered evaluation, standards-based governance, orchestration architectures, and runtime assurance mechanisms. This paper contributes a bounded evidence synthesis across a manually coded corpus of twenty-four recent sources. The core finding is a \emph{governance-to-action closure gap}: evaluation tells us whether outcomes were good, governance defines what should be allowed, but neither identifies where obligations bind to concrete actions or how compliance can later be proven. To close that gap, the paper introduces three linked artifacts: (1) a four-layer framework spanning evaluation, governance, orchestration, and assurance; (2) an \odta{} runtime-placement test based on observability, decidability, timeliness, and attestability; and (3) a minimum action-evidence bundle for state-changing actions. Across sources, evaluation papers identify safety, robustness, and trajectory-level measurement as open gaps; governance frameworks define obligations but omit execution-time control logic; orchestration research positions the control plane as the locus of policy mediation, identity, and telemetry; runtime-governance work shows path-dependent behavior cannot be governed through prompts or static permissions alone; and action-safety studies show text alignment does not reliably transfer to tool actions. A worked enterprise procurement-agent scenario illustrates how these artifacts consolidate existing evidence without introducing new experimental data.
\end{abstract}


\section{Introduction}
Agentic AI systems differ from single-turn generative systems because they plan, invoke tools, preserve intermediate state, coordinate with external services, and generate trajectories whose effects accumulate over time. This changes the central question from ``did the model answer correctly?'' to ``did the system behave acceptably throughout execution?'' The urgency of that shift is visible in current capability tracking: Stanford HAI's 2026 AI Index summary reports that Terminal-Bench success on real-world tasks rose from 20\% in 2025 to 77.3\% in 2026 while emphasizing that capability gains are outpacing our ability to measure and manage AI systems \cite{aiindex2026}.

Recent survey work on LLM-agent evaluation argues that current assessment remains especially weak on safety, robustness, cost-efficiency, and scalable fine-grained measurement \cite{survey_eval_agents}. Benchmark-methodology work strengthens the warning: the Agentic Benchmark Checklist (ABC) shows that seemingly small benchmark design choices can under- or overestimate agent performance by up to 100\% in relative terms \cite{abc}. Meanwhile, evaluation is broadening in scope. ``General Agent Evaluation'' argues that hidden domain-specific assumptions distort claims about generality \cite{general_agent_eval}, and MultiAgentBench measures not only task completion but also collaboration and competition quality through milestone-based indicators \cite{multiagentbench}. The direction is clear: endpoint success remains necessary, but it is no longer enough.

At the same time, governance and standardization are accelerating. ISO/IEC 42001, ISO/IEC 23894, ISO/IEC 42005, ISO/IEC 5338, and ISO/IEC 38507 provide organizational baselines for AI management, risk, impact assessment, lifecycle discipline, and governance implications \cite{iso42001,iso23894,iso42005,iso5338,iso38507}. NIST's AI Risk Management Framework (AI RMF 1.0), its Generative AI Profile, and the 2026 AI Agent Standards Initiative reinforce the point that trustworthy AI increasingly requires explicit treatment of risk, control, identity, interoperability, and evidence \cite{nistairmf,nistgaiprofile,nistagentinitiative}. Yet these documents deliberately stop short of specifying a complete execution-time control architecture for agents.

In parallel, orchestration research increasingly treats planning, policy enforcement, state management, and quality operations as elements of a coherent orchestration layer rather than as ad hoc middleware \cite{orchestration_survey}. Runtime-governance work goes further by arguing that the execution path itself is the relevant object of control, because prompts and static permissions are insufficient for non-deterministic, path-dependent systems \cite{policies_on_paths,trace_assurance}. Action-safety work reaches a similar conclusion from another angle: current agents remain weak on safety benchmarks \cite{agent_safetybench}; text-level refusal does not reliably transfer to tool-call safety \cite{mindgap}; and claims about deployed guardrails may themselves require attestation \cite{proofguardrail}.

Taken together, these literatures point to a synthesis problem. Agentic AI is no longer only a capability problem, nor only a standards problem, nor only an architecture problem. It is a cross-layer control problem. The recurring mismatch between what governance says must hold and the point in the system where a concrete action can actually be blocked, rewritten, approved, or later proven is named in this paper the \emph{governance-to-action closure gap}.

This paper addresses that gap through a bounded evidence synthesis and four linked contributions:
\begin{enumerate}[leftmargin=*, itemsep=0.15em]
    \item It synthesizes a manually coded corpus of twenty-four recent sources spanning evaluation, governance, orchestration, action-safety, and assurance.
    \item It introduces the governance-to-action closure gap as the paper's core synthesis finding.
    \item It proposes a four-layer framework and a reusable \odta{} runtime-placement test for assigning requirements to design time, runtime mediation, assurance, or human oversight.
    \item It derives a minimum action-evidence bundle (\maeb{}) and demonstrates the framework through a worked enterprise procurement-agent scenario.
\end{enumerate}

The central thesis is that trustworthy agentic AI cannot be assessed through endpoint success alone. Reliable evaluation must incorporate governance objectives and orchestration-level enforcement, because safety and trust emerge from controlled execution paths and evidentiary traces, not merely from final outputs. The novelty is therefore not another isolated guardrail proposal, but a bridge artifact that explains how obligations become action-time checks and how those checks later become evidence.

\section{Methodological Protocol and Corpus Construction}
This work is a \emph{bounded evidence synthesis}, not a pooled meta-analysis of benchmark scores, and it does not claim PRISMA-style exhaustiveness. That distinction matters because current agent benchmarks vary widely in task assumptions, reward design, integration requirements, and measurement units, and recent methodology work shows that such differences can materially distort conclusions \cite{abc}. The goal here is therefore not to compute a single aggregate performance number, but to integrate the strongest available evidence into a coherent analytical framework.

Corpus construction followed a transparent, manually bounded protocol. On Apr. 17, 2026, targeted retrieval was conducted across arXiv, ACL Anthology, Proceedings of Machine Learning Research, NIST, and ISO using eight query families: ``agent evaluation,'' ``agent benchmark methodology,'' ``general agent evaluation,'' ``multi-agent benchmark,'' ``runtime governance for AI agents,'' ``agent orchestration policy enforcement,'' ``tool safety for LLM agents,'' and ``guardrail attestation for AI agents.'' Official ISO and NIST artifacts were then added as governance baselines, and Stanford HAI's 2026 AI Index summary was added as a capability-tracking anchor \cite{aiindex2026}. The retained corpus contains twenty-four sources: fifteen research papers and nine official framework or standards artifacts.

Source selection was role-based rather than exhaustive. A source was retained only if it materially informed at least one of four synthesis questions: \emph{what is being measured}, \emph{what obligations are defined}, \emph{where control is applied}, or \emph{what evidence later proves compliant behavior}. Purely single-turn LLM papers, generic ethics commentary without operational relevance, and materials that did not speak to multi-step action, tool use, or governance were excluded. This means the paper is intentionally narrower than a broad survey, but stronger on bridgeability across streams.

Each retained source was manually coded along eight dimensions: (i) literature stream, (ii) agent setting, (iii) primary unit of analysis, (iv) main quality target, (v) control locus, (vi) evidence type, (vii) dominant failure mode, and (viii) implied enforceability class. Synthesis then proceeded in three steps. First, each source was summarized within its native stream in terms of what it measures, governs, or operationalizes. Second, sources were contrasted across streams to identify recurring mismatches between obligations, measurements, execution-time control loci, and evidence. Third, those mismatches were consolidated into the governance-to-action closure gap, the four-layer framework, the \odta{} test, and the \maeb{} artifact. The framework is therefore not presented as intuition dressed as taxonomy; it is the artifact produced by cross-stream comparison.

\section{What the Evidence Already Shows}
\subsection{Evaluation Has Moved Beyond Endpoint Success, but Not Far Enough}
The evaluation literature now clearly recognizes that agent assessment must extend beyond simple task completion. The 2025 survey on LLM-agent evaluation organizes the field around capabilities, application settings, generalist agents, and evaluation frameworks, and identifies unresolved gaps in safety, robustness, cost-efficiency, and fine-grained scalable methods \cite{survey_eval_agents}. ABC sharpens the methodological critique by showing that benchmark construction choices can substantially distort reported agent performance \cite{abc}. The implication is not merely that agents are hard to benchmark, but that benchmark numbers are meaningful only when their control assumptions are explicit.

Two newer lines strengthen the same point. ``General Agent Evaluation'' argues that fair comparison of general-purpose agents requires a unified protocol that does not bake domain-specific glue into each benchmark \cite{general_agent_eval}. MultiAgentBench extends evaluation beyond task completion by introducing milestone-based measures of collaboration and competition quality in multi-agent settings \cite{multiagentbench}. Together, these papers show that evaluation is shifting from final-answer scoring toward system-level behavior, but still lacks a strong connection to control placement and post hoc evidence.

\subsection{Governance Frameworks Define Obligations, Not Execution Logic}
Standards and public governance frameworks increasingly provide a mature baseline for trustworthy AI. ISO/IEC 42001 addresses AI management systems; ISO/IEC 23894 provides AI-specific risk management guidance; ISO/IEC 42005 structures AI impact assessment; ISO/IEC 5338 covers AI lifecycle processes; ISO/IEC 38507 addresses organizational governance implications; and NIST's AI RMF and Generative AI Profile organize risk management across design, development, deployment, use, and evaluation \cite{iso42001,iso23894,iso42005,iso5338,iso38507,nistairmf,nistgaiprofile}. NIST's AI Agent Standards Initiative extends that momentum into agent-specific identity, authorization, and protocol concerns \cite{nistagentinitiative}.

These artifacts are foundational but intentionally high-level. They clarify what should be governed, monitored, and improved; they do not directly answer where an approval threshold should be checked, how an unsafe tool invocation should be blocked, or what evidence must be captured to replay a disputed action. Their practical value in agentic systems therefore depends on translation into technical controls.

\subsection{Orchestration Is Becoming the Action-Time Control Plane}
The orchestration literature identifies where that translation can happen. The 2026 survey on orchestrated multi-agent systems formalizes a unified orchestration layer that integrates planning, policy enforcement, state management, and quality operations \cite{orchestration_survey}. This matters because it shifts the discussion from vague ``guardrails'' toward a specific architectural locus where controls can be attached: routing, scoped identities, tool permissions, approval gates, state mediation, observability, and rollback.

This also helps reconcile standards with implementation. Governance tells organizations what must be constrained or evidenced; orchestration helps determine where those constraints can realistically be applied while the system is running.

\subsection{Path-Dependent Governance and Assurance Need More than Prompts}
Recent runtime-governance work argues that the central object of control is not the prompt, but the execution path. ``Runtime Governance for AI Agents: Policies on Paths'' formalizes compliance policies as deterministic functions over agent identity, partial path, proposed next action, and organizational state \cite{policies_on_paths}. ``A Trace-Based Assurance Framework for Agentic AI Orchestration'' makes a complementary move by instrumenting executions as message-action traces with step and trace contracts, stress testing, fault injection, and governance outcome distributions \cite{trace_assurance}. Together, these papers show why prompt-level rules and static access control are insufficient for path-dependent systems with external side effects.

\subsection{Action-Safety and Verifiable Guardrails Expose the Hardest Gaps}
Action-safety work supports the same conclusion from another angle. Agent-SafetyBench evaluates 16 popular agents across 349 environments and 2,000 test cases and reports that none exceeds a safety score of 60\% \cite{agent_safetybench}. ToolSafety, an EMNLP 2025 dataset, shows that safety failures in tool-using agents are especially acute in multi-step and indirect-harm settings \cite{toolsafety}. WebGuard shows that frontier models remain below 60\% accuracy in predicting web-action outcomes and below 60\% recall on high-risk actions without dedicated safeguards \cite{webguard}. Mind the GAP then demonstrates a structural failure mode: text-level refusal does not reliably transfer to tool-call safety \cite{mindgap}.

Control papers sharpen the picture further. ToolSafe reports that proactive step-level intervention can reduce harmful tool invocations of ReAct-style agents by 65\% on average while improving benign task completion under prompt-injection attacks \cite{toolsafe}. ShieldAgent proposes explicit safety-policy reasoning over action trajectories and reports 90.1\% recall on its benchmark while also reducing API queries and inference time \cite{shieldagent}. AgentDoG argues that agent safety needs trajectory-aware diagnostics and root-cause visibility rather than binary moderation alone \cite{agentdog}. Proof-of-Guardrail highlights an assurance problem that benchmark scores often ignore: if a guardrail is claimed, users may need evidence that the guardrail actually executed before the response or action was produced \cite{proofguardrail}.

Across streams, the same mismatch recurs. Evaluation tells us whether outcomes look acceptable. Governance tells us which obligations must hold. Orchestration determines where live mediation occurs. Assurance determines whether that mediation can later be demonstrated. The literature already contains each piece, but not a stable way to reason across them. That recurring mismatch is the governance-to-action closure gap.

\section{From Closure Gap to Framework}
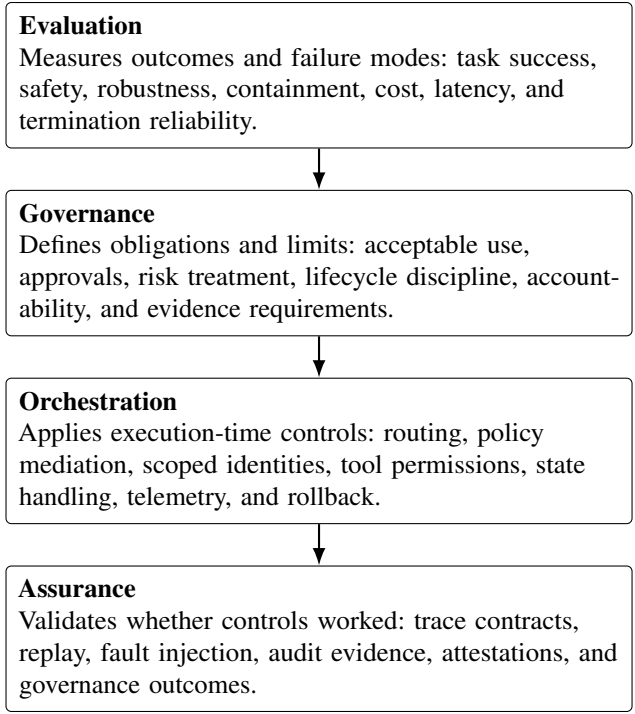
\begin{figure}[t]
\centering
\begin{tikzpicture}[
node distance=0.55cm,
box/.style={draw, rounded corners=2pt, align=left, inner sep=5pt, text width=0.9\columnwidth, minimum height=0.9cm},
arrow/.style={-Latex, thick}
]
\node[box] (eval) {\textbf{Evaluation}\\Measures outcomes and failure modes: task success, safety, robustness, containment, cost, latency, and termination reliability.};
\node[box, below=of eval] (gov) {\textbf{Governance}\\Defines obligations and limits: acceptable use, approvals, risk treatment, lifecycle discipline, accountability, and evidence requirements.};
\node[box, below=of gov] (orch) {\textbf{Orchestration}\\Applies execution-time controls: routing, policy mediation, scoped identities, tool permissions, state handling, telemetry, and rollback.};
\node[box, below=of orch] (assure) {\textbf{Assurance}\\Validates whether controls worked: trace contracts, replay, fault injection, audit evidence, attestations, and governance outcomes.};
\draw[arrow] (eval) -- (gov);
\draw[arrow] (gov) -- (orch);
\draw[arrow] (orch) -- (assure);
\end{tikzpicture}
\caption{The Beyond Task Success framework. The governance-to-action closure gap sits between normative obligations and concrete action-time mediation; orchestration and assurance are what make that gap tractable rather than rhetorical.}
\label{fig:framework}
\end{figure}

The closure gap motivates the paper's central artifact: a four-layer framework shown in Fig.~\ref{fig:framework}. The layers are complementary rather than substitutable. They answer different questions, and trust breaks when those questions are collapsed into one. In compact form, the layers answer: evaluation = what happened, governance = what should happen, orchestration = what may happen now, and assurance = how that claim can later be proven.

\subsection{Layer 1: Evaluation}
Evaluation measures what the system achieves and how it fails. For agents, this cannot stop at final task completion. It must include trajectory-aware properties such as safety, robustness, containment, non-termination, and often cost-efficiency \cite{survey_eval_agents,abc,agent_safetybench,multiagentbench}. Benchmark hygiene is part of evaluation, not a side issue, because flawed task or reward design can manufacture false confidence \cite{abc}.

\subsection{Layer 2: Governance}
Governance defines obligations, accountability, escalation rules, and risk treatment. This layer answers questions such as which actions require human approval, which resources are in scope, what evidence must be retained, and which harms require formal impact assessment \cite{iso42001,iso23894,iso42005,iso5338,iso38507,nistairmf,nistgaiprofile}. Governance is normative: it describes what should hold, but not by itself how live systems enforce it.

\subsection{Layer 3: Orchestration}
Orchestration is the execution-time control plane. It is where routing, delegated identities, memory and state, tool permissions, approvals, and telemetry are coordinated \cite{orchestration_survey}. This is the layer most often mistaken for ``the'' guardrail layer. In reality, it is the mechanism that operationalizes only a subset of governance requirements: the subset that can be judged from available runtime context.

\subsection{Layer 4: Assurance}
Assurance validates whether controls actually worked. Trace contracts, deterministic replay, structured fault injection, attestations, and governance outcome distributions belong here \cite{trace_assurance,proofguardrail}. Assurance is what keeps the paper from collapsing into a mere architecture proposal. Without assurance, orchestration may enforce rules, but there is no disciplined way to determine whether the right rules fired, where the first violation occurred, or whether the system remained within acceptable bounds.

\section{The \odta{} Runtime-Placement Test}
A central contribution of this paper is the claim that not every requirement belongs at runtime. To operationalize that claim, the synthesis yields a four-question test based on \textbf{O}bservability, \textbf{D}ecidability, \textbf{T}imeliness, and \textbf{A}ttestability (\odta{}):
\begin{enumerate}[leftmargin=*, itemsep=0.12em]
    \item \textbf{Observability}: Can the protected event be observed before or at the moment of action?
    \item \textbf{Decidability}: Is the rule crisp enough for deterministic or probabilistic automation without unacceptable ambiguity?
    \item \textbf{Timeliness}: Can intervention happen early enough to reduce risk without defeating the task itself?
    \item \textbf{Attestability}: If enforcement is imperfect, can the relevant evidence still be reconstructed from logs, traces, signatures, or attestations?
\end{enumerate}

The test is intentionally simple. Its purpose is not to replace domain judgment, but to prevent a common failure mode in agent governance discussions: assuming that because a requirement matters, it must therefore be enforceable at runtime. In practice, the placement logic is:
\begin{equation}
\text{runtime candidate}(r) \Leftarrow O(r) \wedge D(r) \wedge T(r)
\end{equation}
\begin{equation}
\text{assurance-first}(r) \Leftarrow A(r) \wedge \neg\bigl(O(r) \wedge D(r) \wedge T(r)\bigr)
\end{equation}
where $O, D, T, A \in \{0,1\}$ indicate whether a requirement satisfies the four \odta{} properties strongly enough for the deployment context. Human-judgment dependence typically appears when decidability is low because compliance turns on contested values, context, or organizational interpretation. Design-time placement dominates when the protected event is structurally unavailable at runtime and cannot be reconstructed later.

\begin{table}[t]
\caption{The \odta{} runtime-placement test}
\label{tab:odta}
\centering
\footnotesize
\begin{tabularx}{\columnwidth}{@{}>{\raggedright\arraybackslash}p{0.22\columnwidth}>{\raggedright\arraybackslash}X>{\raggedright\arraybackslash}p{0.24\columnwidth}@{}}
\toprule
Class & Diagnostic reading of \odta{} & Typical control locus \\
\midrule
Design-time only & Key runtime facts are not observable or reconstructable; control depends mainly on architecture, process, vendor choice, or lifecycle design. & Governance, architecture, lifecycle process \\
Runtime-enforceable & Observability, decidability, and timeliness are all strong enough to block, rewrite, or require approval before the action commits. & Orchestrator, policy engine, approval gate \\
Post hoc auditable & Full runtime prevention is weak or partial, but sufficient evidence can still reconstruct whether policy held. & Assurance, audit, replay workflow \\
Human-judgment dependent & Rule semantics are contested, value-laden, or too context-dependent to reduce safely to deterministic mediation. & Governance board, human review, exception handling \\
\bottomrule
\end{tabularx}
\end{table}

The \odta{} test also yields a stronger negative result: requirements with low decidability or low timeliness are not failed runtime rules; they are often \emph{misplaced} runtime rules. That distinction matters because over-aggressive mediation can reduce mission success without meaningfully improving trust.

\section{Coverage Matrix Across Existing Evidence}
The synthesized evidence suggests that current artifacts cover the problem unevenly. Table~\ref{tab:coverage} highlights where representative sources are strongest and where gaps remain. The matrix is illustrative rather than exhaustive; its purpose is to show which streams speak most directly to outcomes, path control, live enforcement, and assurance evidence.

\begin{table*}[t]
\caption{Illustrative coverage matrix across representative sources}
\label{tab:coverage}
\centering
\footnotesize
\begin{tabularx}{\textwidth}{@{}>{\raggedright\arraybackslash}p{0.23\textwidth}>{\centering\arraybackslash}p{0.07\textwidth}>{\centering\arraybackslash}p{0.08\textwidth}>{\centering\arraybackslash}p{0.09\textwidth}>{\centering\arraybackslash}p{0.10\textwidth}>{\raggedright\arraybackslash}X@{}}
\toprule
Source family & Outcomes & Path view & Runtime control & Assurance & Main contribution \\
\midrule
Evaluation surveys and benchmark-rigor work \cite{survey_eval_agents,abc} & High & Medium & Low & Low & Map the field and expose where benchmark design distorts conclusions \\
General and multi-agent evaluation \cite{general_agent_eval,multiagentbench} & High & Medium & Low & Medium & Shift evaluation toward generality, coordination, and milestone-based behavior \\
Action-safety benchmarks \cite{agent_safetybench,toolsafety,webguard,mindgap} & Medium & High & Low & Low & Show that tool-use and action risk are not captured by text-only safety \\
Control and guardrail papers \cite{toolsafe,shieldagent,agentdog,proofguardrail} & Medium & High & High & High & Show that intervention, diagnostics, and guardrail evidence are separate design problems \\
Runtime-governance and trace-based assurance \cite{policies_on_paths,trace_assurance} & Low & High & High & High & Formalize path-dependent governance and trace contracts \\
Orchestration research \cite{orchestration_survey} & Low & Medium & High & Medium & Locates policy, state, identity, and quality operations in the orchestration layer \\
ISO/NIST frameworks \cite{iso42001,iso23894,iso42005,iso5338,iso38507,nistairmf,nistgaiprofile,nistagentinitiative} & Low & Low & Low & Medium & Define obligations, accountability, and trust objectives rather than execution logic \\
\bottomrule
\end{tabularx}
\end{table*}

The matrix reveals the paper's core synthesis result. No single stream covers the full stack. Evaluation work is strongest on outcome measurement and benchmark critique, but usually weak on enforcement logic. Governance frameworks define what matters, but not how live control occurs. Runtime-governance papers and step-level guardrails illuminate the path object, but they do not substitute for broader accountability frameworks. Orchestration research explains where controls can reside, yet often leaves evaluation and assurance underdeveloped. The need for an integrated framework follows directly from this asymmetry.

\section{Derived Artifact: The Minimum Action-Evidence Bundle}
The literature also motivates a more concrete artifact than a framework alone. If a state-changing action cannot be supported by a minimum evidence bundle, then claims of runtime governance remain difficult to falsify, replay, or audit. This paper therefore derives a \emph{minimum action-evidence bundle} (\maeb{}) for externally consequential actions, shown in Table~\ref{tab:maeb}. The bundle is intentionally narrow: it is the minimum evidence required to connect governance intent, runtime mediation, and post hoc assurance.

\begin{table*}[t]
\caption{Minimum action-evidence bundle (\maeb{}) for state-changing actions}
\label{tab:maeb}
\centering
\footnotesize
\begin{tabularx}{\textwidth}{@{}>{\raggedright\arraybackslash}p{0.23\textwidth}>{\raggedright\arraybackslash}p{0.27\textwidth}>{\raggedright\arraybackslash}X@{}}
\toprule
Bundle element & Why it matters & Example artifact \\
\midrule
Delegated identity and authority scope & Prevents actions from appearing authorized when the acting identity or scope was ambiguous & Service account, user delegation token, allowed role scope \\
Policy identifier and decision record & Links the action to the exact policy and decision path that mediated it & Policy version, threshold rule ID, allow/rewrite/block verdict \\
Pre-action state reference & Makes the action replayable against the state actually seen at decision time & Budget snapshot hash, vendor-list version, approval status snapshot \\
Proposed action and resolved parameters & Records what the agent was about to do, not only the user-level request & Tool name, endpoint, arguments, recipient, amount \\
Mediation outcome and exception path & Distinguishes unmediated execution from approved or rewritten execution & Approval token, human override ID, exception code \\
External side-effect handle & Connects the internal decision trace to what happened in the outside system & Purchase-order number, message ID, database transaction ID \\
Trace linkage and optional attestation & Supports replay and, where needed, proof that a guardrail actually executed & Trace ID, signed log pointer, TEE attestation or guardrail proof \\
\bottomrule
\end{tabularx}
\end{table*}

\maeb{} is not a new standard, and it is not meant to replace richer domain-specific audit records. Its value is narrower and more practical: it gives system designers a minimal evidentiary test. If a deployment cannot produce this bundle for a risky state-changing action, then its governance claims are likely under-specified, its orchestration claims are difficult to verify, or its assurance story is incomplete \cite{trace_assurance,proofguardrail}.

\section{Worked Scenario: Enterprise Procurement Agent}
To show how the framework can be applied without new experiments, consider an enterprise procurement agent that can query approved catalogs, retrieve contract data, compare quotes, draft purchase orders, and communicate with human approvers or external vendors.

A stepwise view of the scenario makes the layer boundaries concrete:
\begin{enumerate}[leftmargin=*, itemsep=0.1em]
    \item The agent receives a purchase request and extracts the requester, amount, urgency, and category.
    \item It queries approved vendors, catalog terms, contract constraints, and current budget state.
    \item It computes policy state such as spend threshold, requester role, and whether delegated authority is sufficient.
    \item If the request exceeds a threshold or hits an exception class, it pauses and seeks a human approval token.
    \item Only after policy checks succeed does it issue a purchase order or communicate with a vendor.
    \item It persists the full \maeb{} plus trace linkage for replay, audit, and dispute resolution.
\end{enumerate}

This scenario is useful because it is policy-bounded, tool-using, and path-dependent. It also exposes the difference between endpoint success and acceptable behavior. A procurement agent could complete its task while still violating approval policy, overstepping its access rights, or creating irreproducible side effects.

\begin{table*}[t]
\caption{Applying the framework to a procurement-agent scenario}
\label{tab:scenario}
\centering
\footnotesize
\begin{tabularx}{\textwidth}{@{}>{\raggedright\arraybackslash}p{0.28\textwidth}>{\raggedright\arraybackslash}p{0.17\textwidth}>{\raggedright\arraybackslash}p{0.22\textwidth}>{\raggedright\arraybackslash}X@{}}
\toprule
Requirement & \odta{} outcome & Primary implementation layer(s) & Example evidence or metric \\
\midrule
Only approved vendors may receive purchase orders & Runtime-enforceable (high O/D/T, medium A) & Governance + orchestration & Pre-dispatch allowlist decision and immutable action log \\
Purchases above a defined threshold require human approval & Runtime-enforceable plus assurance & Governance + orchestration + assurance & Approval token, actor identity, timestamp, and policy-firing trace \\
The agent may access only procurement-relevant systems and scopes & Design-time plus runtime-enforceable & Governance + architecture + orchestration & Scoped credentials, denied-action logs, and access telemetry \\
Supplier ranking should remain fair and contestable & Human-judgment dependent with post hoc audit & Governance + assurance + human review & Periodic audit report, explanation template, exception register \\
All state-changing actions must be attributable and replayable & Post hoc auditable plus runtime capture & Orchestration + assurance & End-to-end \maeb{}, trace completeness, and replay success rate \\
\bottomrule
\end{tabularx}
\end{table*}

Several insights follow. First, the strongest runtime candidates are rules whose protected event is observable before execution and whose decision criterion is crisp. Vendor allowlists and approval thresholds are examples. Second, broad requirements such as fairness or contestability are poor candidates for deterministic runtime checks; they belong primarily in governance, review, and assurance. Third, some requirements span layers. Replayability, for example, depends on runtime capture but is judged in assurance.

The scenario also shows why orchestration is the bridge between governance and assurance. Governance specifies the approval threshold and role boundaries; orchestration applies those boundaries during execution; assurance later validates whether the \maeb{} and trace actually demonstrate compliant behavior.

Three failure variants make the point sharper. In an \emph{under-binding} variant, the agent retrieves a stale vendor record and sends a valid-looking purchase order to a non-approved supplier. The task appears complete, but governance and orchestration have failed because the allowlist check did not bind to the dispatch event. In an \emph{evidence-poor} variant, the agent pauses for approval but records the approval token without actor binding, state snapshot, or replay linkage. The business action may be acceptable, yet assurance fails because attribution and replayability are incomplete. In an \emph{over-binding} variant, a coarse policy blocks all off-catalog vendors, including a legitimate emergency purchase that should have followed an exception path. Here safety improves only superficially while mission success degrades. These variants show why the four layers can fail independently and why ``more guardrails'' is not the same thing as better control.

\section{Discussion}
\subsection{Why This Framework Matters}
The main value of the framework is disciplinary and methodological. It prevents four common category errors. First, it prevents treating benchmark success as a proxy for trustworthy behavior. Evaluation literature increasingly rejects that simplification, especially for agents with long horizons and external side effects \cite{survey_eval_agents,abc,agent_safetybench,multiagentbench}. Second, it prevents treating standards as if they were executable policies. ISO and NIST frameworks define governance expectations, but they are not themselves runtime control languages \cite{iso42001,nistairmf}. Third, it prevents treating orchestration as mere plumbing. Orchestration is the place where many live controls actually reside \cite{orchestration_survey}. Fourth, it prevents treating runtime guardrails as sufficient by themselves. Assurance and post hoc evidence remain indispensable \cite{trace_assurance,proofguardrail}.

\subsection{Implications for Benchmark Design}
The synthesis suggests three implications for benchmark design. Benchmarks should measure more than completion, especially containment, intervention quality, and trace-level compliance. They should document orchestration assumptions explicitly, because evaluation outcomes depend on what the control plane mediates or hides. And they should distinguish design-time safety mechanisms from live runtime intervention, since those are different forms of control. These recommendations follow naturally from ABC, Agent-SafetyBench, MultiAgentBench, ToolSafety, and trace-based assurance \cite{abc,agent_safetybench,multiagentbench,toolsafety,trace_assurance}.

\subsection{Implications for Enterprise Deployment}
For practitioners, the framework suggests that the orchestrator should be treated as a policy enforcement plane rather than as a passive scheduler. This is also why the NIST AI Agent Standards Initiative matters: secure interoperability, identity, and protocol design are not peripheral concerns once agents begin acting on behalf of users across systems \cite{nistagentinitiative}. A mature enterprise deployment therefore needs at least: scoped identities, policy mediation, approval logic, step telemetry, and evidence capture sufficient to emit the \maeb{} for risky state-changing actions.

\subsection{What the Framework Does Not Claim}
The framework deliberately does not claim that all governance norms can be compiled into runtime rules, or that orchestration is the only meaningful place to improve agent trust. Some obligations remain fundamentally organizational, legal, or value-laden. Others are best handled through architecture, vendor selection, lifecycle process, or human escalation. The contribution here is therefore not a universal control language. It is a disciplined way to decide which obligations belong where, and what minimum evidence should exist when risky actions are allowed to proceed. In that sense, the paper's novel claim is about alignment between obligation, mediation, and proof, not about any one layer subsuming the others.

\subsection{Testable Propositions for Future Work}
The framework yields propositions that can be checked in later empirical studies. \emph{P1}: benchmark scores that omit orchestration assumptions will be less stable across deployment contexts than trace-based metrics that expose control-plane decisions. \emph{P2}: requirements with high observability, high decidability, and high timeliness will benefit more from runtime mediation than from prompt-level instruction alone. \emph{P3}: deployments that emit a complete \maeb{} for risky actions will show lower dispute resolution cost and higher replay completeness than deployments that log only final outputs. \emph{P4}: assurance metrics such as replay completeness, contract compliance, and governance outcome distributions will better predict deployment robustness than final task success alone. These propositions make the synthesis falsifiable, which is a stronger posture than treating it as a closed conceptual taxonomy.

\section{Threats to Validity and Limitations}
\subsection{Bounded Rather than Exhaustive Review}
This paper uses a bounded corpus rather than an exhaustive review. That design improves interpretability and keeps the synthesis focused on bridge sources, but it also risks omitting relevant concurrent work. The framework should therefore be read as a disciplined integration of representative streams, not as a final census of the field.

\subsection{Source Maturity and Benchmark Heterogeneity}
Several important sources in this area are recent preprints, and benchmark designs remain highly heterogeneous \cite{survey_eval_agents,abc,toolsafe,mindgap,webguard,agentdog,policies_on_paths,trace_assurance,proofguardrail}. This is precisely why a pooled score meta-analysis was avoided. The resulting framework is robust to heterogeneity, but the surrounding empirical landscape is still moving quickly.

\subsection{Standards Access and Interpretation}
The ISO standards referenced here are publicly described primarily through official scope pages and summaries rather than through exhaustive clause-by-clause analysis, because the complete standards texts are not openly accessible \cite{iso42001,iso23894,iso42005,iso5338,iso38507}. The standards discussion is therefore best understood as standards-informed rather than as a formal compliance crosswalk.

\subsection{Framework Subjectivity}
Any cross-layer synthesis includes interpretive judgment. The assignment of a requirement to governance, orchestration, assurance, or human review may vary by domain, risk tolerance, and system architecture. The \odta{} test reduces arbitrariness, but it does not eliminate it. Future empirical work should test inter-rater consistency for the test and the \maeb{} across additional domains.

\section{Conclusion}
Agentic AI has exposed the limits of judging systems by task success alone. The literature now shows, from multiple directions, that agents must be evaluated for trajectory-level behavior, governed through explicit obligations and accountability structures, controlled through an orchestration layer that mediates live actions, and validated through assurance mechanisms that inspect traces and evidence over time. Yet these streams are often discussed in isolation.

This paper's contribution is to connect them through a single synthesis. The governance-to-action closure gap names the recurring mismatch between normative requirements and action-time control. The four-layer model clarifies how evaluation, governance, orchestration, and assurance answer different but interdependent questions. The \odta{} test explains why some obligations belong at runtime while others belong in design, audit, or human oversight. The \maeb{} makes the assurance requirement concrete by specifying the minimum evidence bundle for risky state-changing actions. The procurement-agent scenario demonstrates how these ideas can be applied without collecting new benchmark data, while the action-safety and guardrail literature shows that control failure can arise even when text outputs look aligned. The central conclusion is straightforward: beyond task success, trustworthy agentic AI is a cross-layer control and evidence problem.

\balance

\end{document}